\documentclass[aps,pra,twocolumn,showpacs,preprintnumbers]{revtex4}

\usepackage{mathbbol}
\usepackage{bbm}
\usepackage{graphicx}

\usepackage[intlimits]{amsmath}
\usepackage{amssymb} 
\usepackage{latexsym}
\usepackage{amsfonts}

\newcommand{\trz}[2]{\, \mathrm{tr}_{#1}\! \left( #2 \right)}

\newcommand{\etal}{{\it et al.\ }}

\newcommand{\ket}[1]{\left| #1\right>}

\newcommand{\ketbra}[2]{\big| {#1}\big> \big< {#2}  \big|}
\newcommand{\proj}[1]{\ketbra{#1}{#1}}

\begin{document}

\title{Experimental generation of pseudo bound entanglement} 
\author{Hermann Kampermann}
\email{kampermann@thphy.uni-duesseldorf.de}
\affiliation{Institut f\"ur Theoretische Physik III, Heinrich-Heine-Universit\"at D\"usseldorf, D-40225 D\"usseldorf, Germany}
\author{Xinhua Peng}
\affiliation{Experimentelle Physik IIIa, Technische Universit\"at Dortmund, D-44221 Dortmund, Germany}
\affiliation{Hefei National Laboratory for Physical Sciences at Microscale and
Department of Modern Physics,
University of Science and Technology of China, Hefei, Anhui 230026, People's Republic of China}
\author{Dagmar Bru\ss}
\affiliation{Institut f\"ur Theoretische Physik III, Heinrich-Heine-Universit\"at D\"usseldorf, D-40225 D\"usseldorf, Germany}
\author{Dieter Suter}
\affiliation{Experimentelle Physik IIIa, Technische Universit\"at Dortmund, D-44221 Dortmund, Germany}

\date{\today}

\begin{abstract}
We use Nuclear Magnetic Resonance (NMR) to experimentally generate
a bound entangled (more precisely: pseudo bound entangled)
state,
i.e.\ a quantum state which is non-distillable but nevertheless
entangled.
Our quantum system consists of three qubits.
We characterize the produced state via state tomography to show that
the created state has a positive partial
transposition with respect to any bipartite splitting, and we use a witness operator to prove its entanglement.

\end{abstract}

\maketitle


Quantum entanglement has been a source for fascination
and surprise
 for more than 70 years \cite{Schroedinger}.
The discovery that entanglement can serve as a resource for
information processing has triggered the development of
the broad field of quantum information science \cite{leuchs,SuterBuch}.
Designing and controlling entangled quantum states has thus become
a major experimental challenge.
Several types of
entangled states have already been experimentally generated and
detected in the recent years, e.g.
bipartite entanglement \cite{BellAspect,TwoIonEntanglement},
tripartite entanglement \cite{PhotonGHZPRL99,IonGHZWineland,IonGHZBlatt},
as well as multipartite
entanglement \cite{PhotonWitnessPRL04,6photonEnt,MultiEntIonBlatt,MultiEntIonWineland}.

A particularly interesting class of entangled states are
``bound entangled'' states \cite{BoundEntH3PRL98}, which carry quantum correlations
of an especially elusive and fragile type.
The name ``bound entanglement'' implies an
 analogy to bound energy in thermodynamics, which cannot
be freed to perform work.
Bound entanglement
can be
described as inherent quantum correlations, without the typical
information processing potential of ``free'' entanglement.
Namely, some correlations of quantum nature are
initially established during the generation of a bound
entangled state, but
nevertheless it is not possible to extract a maximally entangled state
for any number of copies of the state (i.e., entanglement distillation
is not possible for bound entangled states).
However, it has
been shown that under certain circumstances bound entanglement
can nevertheless be useful to establish a secret key
\cite{BoundEntKey}. Bound entanglement is expected to exist
in Nature - for example it was shown theoretically
in \cite{BoundEntSpin} that
thermal states of several spin models carry bound entanglement.
Bound entangled states are always mixed, and due to the small
region that they occupy in the space of all states, they are
vulnerable to decoherence. Therefore, it is challenging to
create and detect them in the laboratory.

Here, we report on the first experiment in
which pseudo bound entanglement is created and observed \cite{TalkCortinaDPG}.
We generate a bound
entangled state of a class initially suggested by
Acin \etal \cite{08ClassThreeQubitEntBruss}
with Nuclear Magnetic Resonance (NMR),
and show their undistillability via state tomography,
by proving that the partial transposes
are positive.  A
suitable entanglement witness is
implemented to detect the entanglement \cite{08HyllusBEGenOWittnessPRA04}.
As usual in room temperature NMR we work with states close to
the (normalised) identity, i.e.\ we consider a state of the form
\begin{equation}\label{Eq-BEppGeneral}
\rho=\frac{1-p}{d}\mathbbm{1}+p\rho_{BE},
\end{equation}
where $\rho_{BE}$ is the bound entangled density matrix.
We therefore call the whole state  $\rho$
pseudo bound entangled. This is
 in analogy with the NMR GHZ state \cite{NMRGHZLaflamme98}.
NMR technology has been widely used to demonstrate quantum
information primitives \cite{NMRalgo,NMRcloning,NMRReviewSuter08,NMRReviewLaflamme08,NMRReviewChuang04},
and here it  proves again to be a versatile tool.

\begin{figure*}
\scalebox{0.8}{\includegraphics{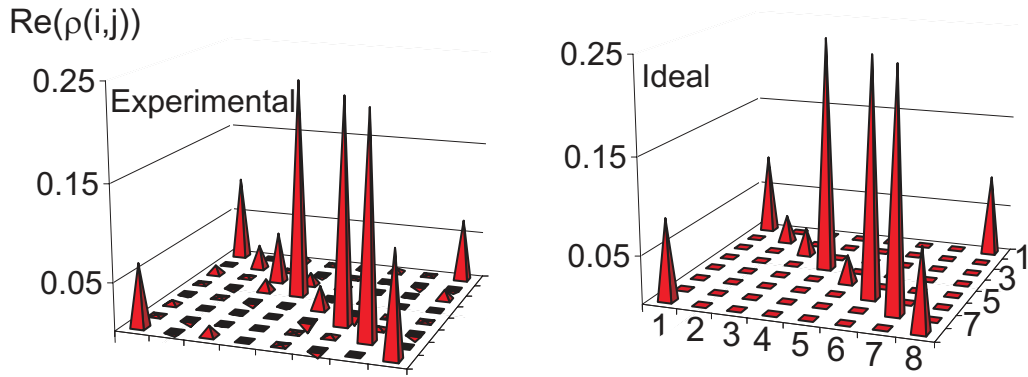}}
\caption{\label{Fig-BERekonstruiert}
Experimental and ideal real matrix elements of $\rho_{\mathrm{BE}}$.
All imaginary elements of the experimental state are small ($<0.019$).
}
\end{figure*}

We consider the family of 
three-qubit
 states 
defined in
\cite{08ClassThreeQubitEntBruss}:
\begin{equation}\label{Eq-rhoBE} \begin{split}
&\rho_{\mathrm{BE}} = 
N^{-1}
\bigg(2 \proj{\mathrm{GHZ}}+\\
&
a_1\proj{001}+a_2\proj{010}+\frac{1}{a_3}\proj{011}+\\
&
\left. a_3\proj{100}+\frac{1}{a_2}\proj{101}+\frac{1}{a_1}\proj{110}\right),
\end{split}
\end{equation}
where 
$\ket{\mathrm{GHZ}}=\frac{1}{\sqrt 2}\left(\ket{000}+\ket{111}\right)$,
and the normalisation is
$N = \left(2+\sum_{i=1}^3( a_i+\frac{1}{a_i})\right)$.
The free parameters
$a_1,a_2,a_3$ are real positive numbers that obey the condition
$a_1a_2a_3\neq 1$.
This familiy of states is ``almost'' diagonal, with the only off-diagonal
matrix elements coming from the GHZ contribution, see Fig. \ref{Fig-BERekonstruiert}
for a visual representation. The states have rank 7 and are thus far from
being pure.
The states in equation (\ref{Eq-rhoBE}) have a positive partial transposition (PPT)
with respect to any bipartite splitting, but it was shown that they
are nevertheless entangled \cite{08ClassThreeQubitEntBruss}.
Therefore, they are  bound entangled.

In \cite{08HyllusBEGenOWittnessPRA04} a family of
witness operators which detects the bound entangled states
in equation (\ref{Eq-rhoBE}) was presented. The family of witnesses has the form
\begin{equation}
\label{eq:witness}
W=\overline W -\varepsilon \mathbbm{1}
\end{equation}
with 
\begin{equation}
\begin{split}
\overline W =& \
\proj{GHZ^-} \\
&
+\frac{1}{1+a_1^2}\bigg(\proj{001}+a_1^2\proj{110}\bigg)\\
& +\frac{1}{1+a_2^2}\bigg(\proj{010}+a_2^2\proj{101}\bigg)\\
& + \frac{1}{1+a_3^2}\bigg(\proj{100}+a_3^2\proj{011}\bigg) \\
&
-\sum_{i=1}^3\frac{a_i}{1+a_i^2}\bigg(\ketbra{000}{111}+\ketbra{111}{000}\bigg).
\end{split}
\end{equation}
where
$\ket{\mathrm{GHZ^-}}=\frac{1}{\sqrt 2}\left(\ket{000}-\ket{111}\right)$.
Note that we are not using any specific normalisation of the witness.
We choose the set of free
parameters for the state and witness in such a way that the witness 
detects entanglement with the highest fraction of white noise.
The optimal parameters calculated via convex optimization
are $a_1=a_2=a_3=a=0.3460$ and $\varepsilon = 0.1069$ \cite{08HyllusBEGenOWittnessPRA04}.

A perfect experimental
realisation of  the  bound entangled state $\rho_{\mathrm{BE}}$
would lead to the expectation value
$
\trz{}{W\rho_{\mathrm{BE}}}=-\varepsilon.
$
This is obvious by construction, as
 $\overline W $ is decomposable
(i.e. it can be written as the sum of a positive operator and the partial
transpose of another positive operator), and therefore its
expectation value
for a PPT bound entangled state cannot be smaller than zero.
In fact, $
\trz{}{\overline W\rho_{\mathrm{BE}}}=0.
$
The expectation value for the identity part gives  $-\varepsilon$.
The smallest possible expectation value of the witness
$W$ is achieved by the GHZ state
$\ket{GHZ}=\frac{1}{\sqrt 2}\left( \ket{000}+\ket{111}\right)$,
because this is the eigenvector corresponding to the negative eigenvalue
of the witness ($\lambda =-\frac{3a}{1+a^2}-\varepsilon\approx -1.03$).
Therefore, the minimal expectation value for a bound entangled state
is only about 1/10 of the minimal expectation value
for a free entangled state. This is one of the reasons why the
detection of bound entanglement is experimentally much more
challenging than the detection of free entanglement.

In current liquid NMR experiments a huge ensemble of about $\sim 10^{19}$ copies of the quantum processors are used. The
ensemble in thermal equilibrium follows the Boltzmann distribution
dominated by the spin interaction with the external magnetic field.
The equilibrium density operator is well approximated by
\begin{equation}\label{Eq-GeneralEqDensOp}
\rho_{\mathrm{eq}}= \frac{1}{d}\left( \mathbbm{1}+ \sum_i \kappa_i I_{zi}\right), 
\end{equation}
where $\kappa_i$ is a spin and system specific constant given by $\kappa_{i}=\frac{\hbar B_0 \gamma_{i}}{k T }$ with
 $\gamma_i$ being the gyromagnetic ratio.
$I_{zi}$ denotes
the $z$-component of the angular momentum operator of the
$i$-th nuclear spin and
$d$ is the dimension of the total system.
In usual liquid NMR experiments with a magnetic field of $\sim 12$ T and protons at ambient temperatures (T= 290 K) we have
$\kappa_{\mathrm{H}}\approx 8.4\cdot 10^{-5}$.

The state of interest in our case is
given by eq. (\ref{Eq-BEppGeneral}),
where we want to generate a specific pseudo bound entangled
state
$\rho$, i.e. the
deviation from the normalized identity,
 with the probability $p$ available in our system.
Our 3-qubit system can reach under the above conditions
a fraction of $p$ for the bound entangled state on the order of $\kappa $.
To characterize
the properties of $\rho_{\mathrm{BE}}$ we will do
state tomography to reconstruct the experimentally generated
state $\rho_{\mathrm{exp}}$. Then we will prove
entanglement of the part \ $\rho_{\mathrm{exp, BE}}$.
For this purpose we adopt the usual witness formalism
to the pseudo state case, i.e.
we shift the witness in eq. (\ref{eq:witness})
by the contribution which comes from the identity, namely
we introduce the ``pseudo witness''
\begin{equation}
W_{\mathrm{NMR}}=\frac{W-\frac{(1-p)}{d}\mathbbm{1}}{p} ,
\end{equation}
such that
$\trz{}{W_{\mathrm{NMR}}\rho}=\trz{}{W\rho_\mathrm{BE}}$.

\begin{figure}
\scalebox{0.6}{\includegraphics[13cm,3.5cm][25cm,10cm]{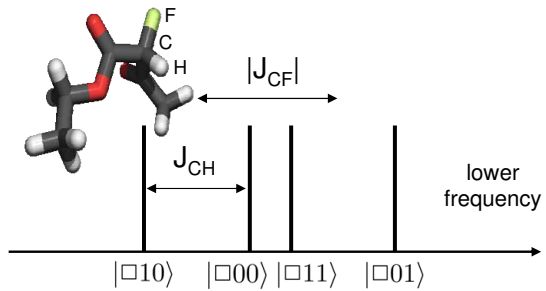}}
\caption{\label{Fig-HCFSpectrum}
Schematic carbon NMR spectrum of the CHF group of
ethyl 2-fluoroacetoacetate and the stick representation of the molecule.
The $J$-coupling between the spins as well as
the H,F spin states in the computational basis are presented. We use the spin state representation of the form
$\ket{\mathrm{C},\mathrm{H},\mathrm{F}}$, i.e.\ C (H,F) corresponds to the first (second, third) qubit, respectively.
The order of the numbering of the spin states in the spectrum is due to the signs
of the $J$-coupling constants. }
\end{figure}

We use the heteronuclear 3-qubit system ethyl 2-fluoroacetoacetate \cite{XinhuaTransferPRA07},
consisting of a hydrogen, a carbon and a fluorine spin (see fig.\ \ref{Fig-HCFSpectrum}) with gyromagnetic ratios of
$\gamma_H \approx 26.75 \cdot 10^7 T^{-1}s^{-1},
\gamma_C \approx  6.73 \cdot 10^7 T^{-1}s^{-1},
\gamma_F \approx 25.18 \cdot 10^7 T^{-1}s^{-1}$.

Using such a heteronuclear system has
several advantages, first we can apply fast local (spin specific) unitary operations, because of the large Zeeman
energy difference between the spins, and second we can focus on the zero order coupling Hamiltonian
in the rotating frame representation,
$H=J_{12}I_{z1}I_{z2}+J_{13}I_{z1}I_{z3}+J_{23}I_{z2}I_{z3}$. Here,
$J_{12}=161.3$ Hz, $J_{13}=-190.2$ Hz and $J_{23}=47.0$ Hz are the j-coupling constants between the spins.

We structure the experiment into three parts:
\begin{itemize}
\item[(i)] Generation of a suitable basis state population.

\item[(ii)]   Unitary transformations (quantum gates) are applied to
             the diagonal state
            to generate $\rho_{\mathrm{BE}}$.

\item[(iii)] The generated state
is characterized by state tomography and witnesses.
\end{itemize}

In the first part of the experiment we
 generate a specific diagonal state, i.e.\
\begin{equation}\label{Eq-InitDiagRho}\begin{split}
\rho_{\mathrm{d}}=& \frac{\mathbbm{1}}{8}+\frac{p}{8}\left( d_a I_{z1}+d_b I_{z2}+d_b I_{z3}+
d_c  I_{z1} I_{z2}+\right.\\
& \left. d_c  I_{z1} I_{z3}+d_d I_{z2} I_{z3}+d_e  I_{z1} I_{z2} I_{z3}\right)
\end{split}\end{equation}
with the coefficients $d_a=\frac{2((a-2)a-1)}{a(3a+2)+3}\approx-0.78$, $d_b=-\frac{2(a+1)^2}{a(3a+2)+3}\approx-0.21$,
$d_c=2 d_a$, $d_d=2 d_b$ and $d_e=\frac{48}{a(3a+2)+3}-8 \approx 3.85$.
We scaled the
traceless spin operators in such a way that the scaling factor $p$ of this deviation density operator corresponds
to the probability of our pseudo bound entangled state of
eq.\ (\ref{Eq-BEppGeneral}).

Five experiments are
performed to generate different basis state populations.
The five diagonal states have the form
\begin{equation}\label{Eq-InitStates}\begin{split}
\rho_1 & =\frac{\mathbbm{1}}{8}+\frac{\kappa_H}{8}\left(3.77\, I_{z1} I_{z2} I_{z3}\right)\\
\rho_2 & =\frac{\mathbbm{1}}{8}+\frac{\kappa_H}{8}\left(-2\, I_{z1} I_{z2}\right)\\
\rho_3 & =\frac{\mathbbm{1}}{8}+\frac{\kappa_H}{8}\left(-1.88\, I_{z1} I_{z3}\right)\\
\rho_4 & =\frac{\mathbbm{1}}{8}+\frac{\kappa_H}{8}\left(-2\, I_{z2} I_{z3}\right)\\
\rho_5 & =\frac{\mathbbm{1}}{8}+\frac{\kappa_H}{8}\left(-I_{z1}  -0.27\, I_{z2} - 0.27\, I_{z3}\right).
\end{split}\end{equation}
The manipulation of the equilibrium states to generate each of these states is done by standard hard spin selective
pulses, j-coupling evolution  \cite{SuterBuch}
and at the end $z$-gradients to destroy the unwanted $x,y$ magnetization terms,
i.e.\ we apply spatial averaging \cite{CoryPhysD}.

The states of these experiments are added with appropriate probabilities to achieve our
 diagonal density operator $\rho_{\mathrm{d}}$ in eq.\ (\ref{Eq-InitDiagRho})
 (temporal averaging \cite{BulkQuantCompStateTom,VandersypenNMREffPureState-OrderFinding}), i.e.\
\begin{equation}\label{Eq-SumTempAv}
\begin{split}
&\rho_{\mathrm{d}}=\sum_i q_i\rho_i\;\;\; \mathrm{with}\;\;\; \sum_i q_i=1\\
& q_1\approx 0.36, q_2=q_5\approx 0.27,
q_3\approx 0.29, q_4\approx 0.08.
\end{split}
\end{equation}
It follows $p\approx\frac{\kappa_H}{ 3.61}$.

On the initial state in eq.\ (\ref{Eq-SumTempAv})
we apply a set of unitary transformations
to produce the pseudo bound entangled state $\rho_{\mathrm{BE}}$. In detail
we first apply an effectively line selective $-\frac{\pi}{2}$-rotation around the $y$-axis in the space
$\ket{000}, \ket{100}$ without affecting the remaining basis states.
In the following set of hard rf-pulses and $j$-coupling evolutions two gates similar to controlled-NOT gates (with qubit 1 as control,
qubit 2 as target and qubit 3 as target, respectively) are applied \cite{NMRGHZLaflamme98}.  The total unitary transformation
of these quantum gates is
\begin{equation}
\begin{split}
U=& \frac{1}{\sqrt 2 }\left(\ketbra{000}{000}+\ketbra{000}{100}-\ketbra{111}{000}+\ketbra{111}{100}\right)\\
& + i\ketbra{001}{001}-i\ketbra{010}{010}+\ketbra{011}{011}\\
& + \ketbra{100}{111}+i\ketbra{101}{110}-i\ketbra{110}{101}.
\end{split}
\end{equation}

The generated state is characterized by state tomography in the usual NMR setting \cite{BulkQuantCompStateTom}, i.e.\
before the detection the spin selective rotations
Y$_1$E$_2$E$_3$, E$_1$E$_2$Y$_3$, E$_1$E$_2$X$_3$, Y$_1$Y$_2$E$_3$, E$_1$X$_2$X$_3$, Y$_1$Y$_2$Y$_3$ and X$_1$X$_2$X$_3$
are performed. Here X$_i$ (Y$_i$) denotes
a $\pi /2$ rotation of the $i$-th spin about the $x$- ($y$-) axis and E$_i$ is the identity operation. In the
experiments only the carbon spin (see fig.\ \ref{Fig-HCFSpectrum}) is detected. To get the information
of the H (F) spins an appropriate swap gate to the carbon and H (F) spin is performed before detection.
The experimental data gained by this procedure leads to an overdetermined set of linear equations for reconstructing the
density matrix. We reconstruct the density operator from the experimental data by a least squares fit \cite{StateTomo}.

The experimentally generated state is shown in Fig.\ \ref{Fig-BERekonstruiert}. The excellent agreement between ideal
and generated state reflects the high controllability as well as low decoherence of our NMR system.
The main deviations are due to
partial relaxation and pulse imperfections during the experiment \footnote{We want to note here that
in the NMR weak measurement scheme the usual statistical errors due to limited number of individual measurements
play no role. I.e.\ it is not necessary to consider reconstruction schemes using maximum likelihood
\cite{LikelihoodHradilLNP04} or Bayesian
mean estimation \cite{BayesMeanBK06}.}.

The fraction $p$ of the pseudo bound entangled
state in $\rho_{\mathrm{exp}}=\frac{(1-p)}{8}\mathbbm{1}+p\rho_{\mathrm{BE}}$
is in our case about $p\approx 10^{-5}$.

How close this state is in comparison to the ideal one
can be expressed in terms of the
overlap or the distance between real and ideal state.
We find a high Uhlmann fidelity \cite{UhlmannFid} for the
pseudo bound entangled state, namely
\begin{equation}
F_u=\trz{}{\sqrt{\sqrt{\rho_{\mathrm{BE,theo}}}\rho_{\mathrm{BE,exp}} \sqrt{\rho_{\mathrm{BE,theo}}}}}=0.98,
\end{equation}
and a small  trace distance \cite{NielsenBookQuantCompInf},
\begin{equation}
d_t=\frac{\trz{}{\sqrt{X^\dag X}}}{2}=0.09, \;\; \mathrm{with}\;\; X=\rho_{\mathrm{BE,theo}}-\rho_{\mathrm{BE,exp}}.
\end{equation}
The standard deviation of the density operator elements are obtained from the least square method. They are used
to calculate the standard deviation of the witness expectation value by error propagation.
This deviation
accounts for the inconsistencies
in the overhead of the experimental data, which are due to imperfections in the tomography procedure.
The reconstructed state $\rho_{\mathrm{BE,exp}}$
has a positive partial transpose with respect to any bipartite splitting.
The witness expectation value of the experimental state is
\begin{equation}
\trz{}{W\rho_{\mathrm{BE,exp}}}=-0.029\pm 0.010,
\end{equation}
i.e.\ the state is bound entangled.
The ideal expectation value would have been $-0.11$. Remember that  the
witness expectation value
can vary from $-1.0$ to $+1.8$,
i.e.\ the possible range
$-0.11\leq \langle W\rangle <0$
for bound entanglement is rather small -
thus, already small deviations from the ideal situation can have large effects.

In conclusion,
we have generated and detected pseudo bound entanglement with an NMR
experiment. We reached a high fidelity, $F=0.98$,  of the created state
with respect to the ideal one.
The partial transpositions of the state were shown
to be positive with respect to any bipartite splitting.
The pseudo witness that we used
had a negative expectation value,
thus proving that the state was indeed entangled.
Here we calculated the expectation value of the witness operator using the reconstructed state. This is more
appropriate than to detect the witness directly \cite{GuehneWitLocalPRA02},
because some experimental errors are
``averaged'' during the least square method used in the state tomography procedure.
- Thus, we have shown that bound entanglement can exist in the laboratory. 

After completion of this 
manuscript we learned about related work by Amselem \textit{et al.\ } \cite{OpticBoundEnt}.

We thank  Otfried G\"uhne and Matthias Kleinmann for enlightening discussions.
We acknowledge financial support from the EU Integrated projects
SCALA and SECOQC, as well as from Deutsche Forschungsgemeinschaft (DFG).

\end{document}